\documentclass[%
 reprint,
prb,
]{revtex4-1}

\usepackage{graphicx}
\usepackage{dcolumn}
\usepackage{bm}
\usepackage{enumitem} 
\usepackage{subfigure}
\usepackage{natbib}
\usepackage{float}
\usepackage{color}
\usepackage{ulem}
\usepackage{amsmath}
\usepackage{ulem}

\newcommand{\COMMENTED}[1]{}
\newcommand{\REMARKS}[1]
{
{ \color{red}{\textbf{ {[#1]} }} }
}

\begin{document}


\title{A Pseudo-BCS Wavefunction from Density Matrix Decomposition: \\
Application in Auxiliary-Field Quantum Monte Carlo}

\author{Zhi-Yu Xiao}
 \affiliation{%
Department  of  Physics,  College  of  William  \&  Mary,  Williamsburg,  Virginia  23187,  USA
}%

\author{Hao Shi}
\affiliation{
Center  for  Computational  Quantum  Physics,  Flatiron  Institute,  New  York,  NY  10010,  USA
}%
\author{Shiwei Zhang}
\affiliation{%
Center  for  Computational  Quantum  Physics,  Flatiron  Institute,  New  York,  NY  10010,  USA
}%
\affiliation{
Department  of  Physics,  College  of  William  \&  Mary,  Williamsburg,  Virginia  23187,  USA
}%
\date{\today}

\begin{abstract}

We present a method to construct pseudo-BCS wave functions from the one-body density matrix. 
The resulting many-body wave function,
which can be produced for any fermion systems, including 
those with purely repulsive interactions, 
has the form of a number-projected BCS form, or 
antisymmetrized germinal power (AGP). 
Such wave functions provide a better ansatz for correlated fermion systems than a single Slater determinant, and often better than a linear combination of Slater determinants (for example from a truncated active space calculation). We describe a procedure to build such a wave function conveniently from a given reduced density matrix of the system,
rather than from a mean-field solution (which gives a Slater determinant for 
repulsive interactions). The pseudo-BCS wave function 
thus obtained reproduces the density matrix or minimizes the difference between the input and resulting 
density matrices.
One application of the pseudo-BCS wave function is in auxiliary-field quantum Monte Carlo (AFQMC) calculations 
as the trial wave function to control the sign/phase problem.
AFQMC 
is often among the most accurate general methods for correlated fermion systems. 
We show that the pseudo-BCS form
further reduces the constraint bias and leads to improved accuracy compared to the usual Slater determinant trial wave functions,
using the two-dimensional Hubbard model as an example. 
Furthermore, the pseudo-BCS trial wave function allows a new 
systematically improvable self-consistent approach, 
with pseudo-BCS trial wave function iteratively generated by AFQMC via the one-body density matrix.
\end{abstract}

\pacs{Valid PACS appear here}

\makeatletter
\newcommand{\rmnum}[1]{\romannumeral #1}
\newcommand{\Rmnum}[1]{\expandafter\@slowromancap\romannumeral #1@}
\makeatother

\maketitle

\section{\label{sec:level1}INTRODUCTION}

The study of strongly-correlated quantum many-body systems is highly challenging. 
A general approach does not yet exist to compress the complexity 
of the many-body wave functions that is widely applicable and yields  systematic accuracy across 
different ranges of many-body models and materials \cite{CPMCModel,CPMCMaterial,simons_material_2020}.
Methodological developments thus have a key role in the study of interacting quantum systems, which  spans several subfields in physics,
including condensed matter physics, nuclear physics, cold atoms physics, as well as in quantum chemistry and materials science. 
Recent work and collaborations on method developments have lead to significant progress with computational approaches. 

The simplest approaches to many-fermion systems are based on the independent-particle framework. The basic entity in this framework is
the Slater determinant. The Slater determinant can be the wave function ansatz itself, as in
a Hartree-Fock (HF) mean-field calculation. Alternatively and more commonly, it is used as a vehicle either to capture some property of the system, for 
example the electronic density and gradients in 
a density-functional theory (DFT) calculations \cite{Kohn_RMP_1999,Martin_book_2004}, or as the starting point or reference state for many-body calculations 
such as perturbation theory (e.g.~MP2) or coupled cluster [e.g.~CCSD(T)] calculations in quantum chemistry \cite{Szabo_book_1989}, or to impose fermion antisymmetry in 
quantum Monte Carlo (QMC) calculations \cite{Foulkes_RMP_2001,
Zhang_Constrained_1997,Zhang-Krakauer-2003-PRL}.
%
The Slater determinant can be manipulated with low-polynomial computational cost while 
fully accounting for permutation antisymmetry, which is a great advantage in treating fermion systems.
The Slater determinant  has a major shortcoming in these roles, however. It is fundamentally a Fermi liquid picture that contains only occupied 
orbitals and discards any information 
on the virtual orbitals. 

 The BCS wave function \cite{BCS,pBCS} is a simple ansatz that allows one to overcome this shortcoming of the Slater determinant, and describe a non-trivial modification to the topology of 
 the independent-electron 
 Fermi surface and momentum distribution. 
 A number-projected BCS wave function can be viewed as a linear combination 
 of Slater determinants.
 It maintains fermion antisymmetry and requires only marginally more computational 
 cost to manipulate.  
This form can therefore be advantageous in various contexts in the study of interacting fermion systems. 
 For example, the BCS form is clearly more desirable in treating systems with superconductivity. 
 Even in more conventional systems, 
 the BCS wave function (usually in its particle-number projected form) has been found to lead to better trial wave functions
 \cite{BCS_in_DMC1,BCS_in_DMC2,Vitali_Calculating_2019, Carlson_2011} 
  in QMC calculations. One can also imagine that a 
 BCS wave function might serve as 
 a better reference state for many-body (e.g., coupled-cluster) calculations.

 The BCS wave function can be generated via a mean-field calculation, 
as  the solution to a Hamiltonian which contains attractive interactions. 
 Conceptually, for a system with  attractive interactions,  
 the BCS wave function can be thought of as a natural way to go beyond the Fermi liquid picture,
 by letting pairs of spin-up and spin-down particles occupy those orbitals which are virtual in 
  the so-called restricted Hartree-Fock (RHF) solution, thereby creating a better 
  variational wave function with lower energy. 
  (For simplicity, we consider a spin-balanced system with singlet pairing only.)
  
  For electronic systems with purely repulsive  Coulomb  interactions
  (such as in most electronic models or atoms, molecules, and most real materials), 
  this picture breaks down.
  If we think of the occupied and virtual orbitals defined by an RHF calculation, 
  the BCS ansatz includes a linear combination of electronic configurations consisting of the RHF and
   all possible paired excitations from it. 
    The pair occupancy of the higher
  virtuals raises the one-particle energy without lowering the interaction energy.
  The lowest energy mean-field  solution is thus
 a Slater determinant, not BCS. 
  
 We consider in this paper a generalized form of the BCS wave function designed for use in electronic systems.
 The pseudo-BCS wave function, as we shall refer to it by, is based on the 
 concept of the one-body
 density matrix instead of the variational energy. 
The wave function is constructed in the structure of a particle-number-projected BCS wave function, with the goal of 
 reproducing the desired
 one-body density matrix of the many-body system. 
 The pseudo-BCS wave function 
 can be thought of as a general pairing form involving the occupied and virtual
 orbitals of, for example, 
an {\it unrestricted\/} Hartree-Fock (UHF)  solution (as opposed to RHF in standard BCS).
As is well known, 
 the UHF determinant, by allowing spin-symmetry-breaking, often provides a better reference description than RHF for
so-called correlated systems (e.g., 
in Hubbard-like models\cite{Shi_Symmetry_2013}, or bond breaking in molecules \cite{AFQMC_Bond_Breaking}). 
In our pseudo-BCS form, we allow the coefficients for orbital pairing occupancy to be complex numbers. The complex phase provides an additional variational degree of freedom 
to lower the interaction energy, whose physical origin will be discussed below.
 We introduce a low-cost 
 algorithm to construct the pseudo-BCS wave function by a decomposition of the density matrix.
 
 Our approach allows a straightforward way to incorporate the pseudo-BCS form into a computational framework beyond mean-field. 
 For example, by coupling it with the
 auxiliary field quantum Monte Carlo (AFQMC) method \cite{Zhang_Constrained_1997,Zhang-Krakauer-2003-PRL}, 
 we obtain a self-consistent procedure in which the pseudo-BCS wave function serves as a trial wave function for the constraint to control 
 the sign/phase problem. The AFQMC calculation computes a one-body density matrix, to which 
 we apply our density matrix decomposition procedure to produce a new pseudo-BCS wave function, and the process is iterated until convergence. 
 We describe the approach below, and illustrate its implementation and performance in the Hubbard model.
 We show that this self-consistent AFQMC can generate a systematically improved trial wave function during the simulation. With the 
 converged pseudo-BCS trial wave function, the AFQMC calculation leads to smaller projection time for reaching the ground state, 
 and smaller systematic bias from the constraint compared with Slater determinant trial wave functions. In the Hubbard model with next-nearest-neighbor
 hopping, the improvement with self-consistent pseudo-BCS trial wave function allows AFQMC to accurately distinguish the subtle spin orders 
 as the hopping amplitudes are varied. 

The rest of this paper is organized as follows.
In Sec.~\ref{sec:level2}, we introduce the formalism of the pseudo-BCS wave function, and show how it can be obtained from the one-body 
density matrix of the many-body state. 
In Sec.~\ref{sec:level3}, we briefly introduce AFQMC and present details to couple 
the AFQMC to pseudo-BCS trial wave functions to achieve self-consistency. 
In Sec.~\ref{sec:level4}, we show illustrative results, and demonstrate the improvement of
the pseudo-BCS wave function and the self-consistent AFQMC with it. 
Then in Sec.~\ref{sec:level5}
we discuss several additional points 
and summarize. 

\section{\label{sec:level2} PSEUDO-BCS WAVEFUNCTION }



The approach discussed in this paper applies generally to any electronic Hamiltonian. To help make the discussion more concrete,
we use the Hubbard model to introduce the pseudo-BCS wave function and the density matrix decomposition algorithm. 
Also the results in Sec.~\ref{sec:level4} will all be for this model. 

The Hubbard Hamiltonian is 
\begin{equation}
H\!\!= -\sum _{\sigma,i,j }\!\!t_{i,j}c^\dagger _{i,\sigma }c_{j,\sigma }+U\sum _{i}n_{i,\uparrow}n_{i,\downarrow}\,,
\label{eq:Ham-Hub}
\end{equation}
where $c^\dagger _{i,\sigma }$ ($c _{i,\sigma }$) creates (annihilates) an electron with spin $\sigma$ ($\sigma = \uparrow, \downarrow$) at lattice site $i$ 
defined in two dimensions on a rectangular lattice of  $N_s\equiv L_x\times L_y$ sites,
and  $n_{i,\sigma} \equiv c^\dagger _{i,\sigma }c_{i,\sigma }$ is the density operator.
Periodic or open boundary condition will be applied.
The hopping matrix elements $\{ t_{i,j}\}$ will contain near-neighbor terms with amplitude $t$ (which is used to set the units of energy) and next-near-neighbor 
terms with amplitude $t'$. 
The parameters $t'/t$ and $U/t$ and the boundary conditions 
will be specified explicitly later for each system. 
In this paper, we use $N_\sigma$ to denote the number of electrons with spin $\sigma$.
As mentioned, we focus in this work on spin-balanced systems, with $N_\uparrow=N_\downarrow$, and no spin-flip terms 
in the Hamiltonian. We comment briefly on generalizations to other cases 
in Sec.~\ref{sec:level5}.

\subsection{\label{sec:level23}Formalism}


In the following, we  introduce the formalism and notations, before discussing how to obtain a pseudo-BCS wave function from a given 
one-body density matrix of the targeted many-body state.
A general Slater determinant, for example the  UHF  solution, can be written as
\begin{equation}
\label{eq:sd}
| \Phi \rangle = \phi^{\dagger}_{1, \uparrow}
 \dots  \phi^{\dagger}_{N_{\uparrow}, \uparrow}\,\phi^{\dagger}_{1, \downarrow} \dots \phi^{\dagger}_{N_{\downarrow}, \downarrow}  \, | 0 \rangle\,,
\end{equation}
where $|0\rangle$ represents the vacuum state and the operator
\begin{equation}
\phi^{\dagger}_{n, \sigma} \equiv \sum_{i} \,  \left(\Phi_{\sigma}\right)_{i,n} \, c^{\dagger}_{i,\sigma}
\end{equation}
creates an electron of spin-$\sigma$ in an orbital described by the $n$-th column vector $(\Phi_{\sigma})_{i,n}$ of the $N_s \times N_\sigma$ matrix 
$\Phi_\sigma$, 
with $\sigma=\uparrow$ or $\downarrow$.
The one-body density matrix of the Slater determinant wave function is given by 
\begin{equation}
\langle c^\dagger_{j,\sigma}c_{i,\sigma} \rangle_\Phi = [\Phi_\sigma (\Phi_\sigma)^\dagger]_{i,j}= \sum_n (\Phi_{\sigma})_{i,n}\,(\Phi_{\sigma})_{j,n}^\star\,. 
\label{eq:DM-det}
\end{equation}

The structure of a particle-number-projected BCS wave function,  also known as anti-symmetrized germinal power (AGP) \cite{AGP_Coleman_1965,AGP_CI_2019}, is 
\begin{equation}
\label{eq:PseudoBCS}
|\Psi\rangle =\underbrace{\psi^\dagger...\psi^\dagger}_{N_{\sigma}}|0\rangle\,,
\end{equation}
where $\psi^\dagger$ is a pair creation operator, defined as  
\begin{equation}
\psi^\dagger \equiv \sum _{i,j}F_{i,j}c^\dagger _{i,\uparrow}c^\dagger _{j,\downarrow}\,.
\end{equation}
The $N_s\times N_s$ matrix $F$ can be written, for example by a singular value decomposition (SVD), in the form 
$F=UDV^\dagger$, with 
\begin{equation}
\label{eq:pairing-SVD}
F_{i,j} =\sum_{n} d_n\,U_{i, n}\,V^\star_{j,n}\,
\end{equation}
where $D$ is a diagonal matrix whose elements are given by $\{d_n\}$, 
and the $N_s\times N_s$ matrices $U$ and $V^\star$ contain single-particle orbitals. 
Our pseudo-BCS wave function will have the same form, and will allow the diagonal elements to be complex:
\begin{equation}
\label{eq:pairing-pBCS}
F_{j,k} = \sum_{n} e^{i\theta_n}|d_n| \,U_{j, n} \, V^\star_{k,n}\,.
\end{equation}
The pseudo-BCS wave function  will couple orbitals described by $U$ and $V^\star$ in a way that represents different 
principles of pairing from standard BCS, as we discuss in the next subsection.

\subsection{\label{sec:Algorithm}Pseudo-BCS wave function from density matrix} 

For the reverse process of Eq.~(\ref{eq:DM-det}), namely to find the best Slater determinant wave function given the density matrix ${\mathcal G}_\sigma$,
one can construct so-called  natural orbitals. The density matrix can be diagonalized to obtain the 
natural orbitals (eigenvectors) and occupancies (eigenvalues). The matrix $\Phi_\sigma$ for the Slater determinant 
is then be formed by the natural orbitals with the 
$N_\sigma$ highest occupancies. Clearly the resulting density matrix from Eq.~(\ref{eq:DM-det}) is an approximation and does not reproduce 
${\mathcal G}_\sigma$ exactly, with information of the higher (virtual) orbitals lost. 

Now we 
consider this process for pseudo-BCS wave functions:
If we have a best estimate of the one-body density matrix ${\mathcal G}_\sigma$,
\begin{equation}
({\mathcal G}_\sigma)_{i,j}=\frac{\langle \Psi_0 |c^\dagger_{j,\sigma}c_{i,\sigma}|\Psi_0 \rangle}{\langle \Psi_0 |\Psi_0 \rangle}\,,
\label{eq:DM-gs}
\end{equation}
where $|\Psi_0\rangle$ representing the 
the many-body ground-state wave function.
 how to obtain a pseudo-BCS wave function which best reproduces the density matrix.
 This is accomplished 
 using a decomposition of the density matrix with no truncation of the occupancy, as we illustrate next.

If ${\mathcal G}_\uparrow$ and ${\mathcal G}_\downarrow$ 
share the same eigenvalues, there exists a pseudo-BCS wave function which can reproduce the density matrix exactly. To demonstrate the procedure, we work with the natural orbitals and occupancy for each spin sector:
\begin{equation}
{\mathcal G}_\uparrow = P \Lambda P^\dagger\,; \quad {\mathcal G}_\downarrow= Q\Lambda Q^\dagger\,,
\label{eq:target_DM}
\end{equation}
where $\Lambda$ is a diagonal matrix with eigenvalues $\{\lambda_1, \lambda_2, \cdots, \lambda_{N_s}\}$, the common occupancy numbers. 
Using the natural orbitals:
 \begin{equation}
\tilde  c^\dagger_{n,\uparrow } = \sum_i P_{i,n} c^\dagger_{i,\uparrow}\,;
\quad  
\tilde  c^\dagger_{n,\downarrow } = \sum_j Q_{j,n} c^\dagger_{j,\downarrow}\,,
\end{equation}
we can write down a pseudo-BCS wave function:
\begin{equation}
\label{eq:BCS-expand-det}
\begin{aligned}
|\Psi\rangle&=\biggl(\sum _{n}d_{n} \tilde c^\dagger _{n,\uparrow}\, \tilde c^\dagger _{n,\downarrow} \biggr )^{N_{\sigma}}|0\rangle \\
&=\sum _{\{n\}} d_{n_1}...d_{n_{N_{\sigma}}}\, \tilde c^\dagger _{n_1,\uparrow} \tilde c^\dagger_{n_1,\downarrow} \cdots \tilde c^\dagger _{n_{N_{\sigma}},\uparrow}\, \tilde c^\dagger _{n_{N_{\sigma}},\downarrow}\,|0 \rangle\,,
\end{aligned}
\end{equation}
where $\{n\}=\{n_1, n_2, \cdots, n_{N_\sigma}\}$ is a set of $N_\sigma$  non-repetitive indices from $\{1,2, \cdots, n, \cdots, N_s\}$. 

In the basis of these natural orbitals, the density matrix 
of $|\Psi\rangle $ is
\begin{equation}
(\tilde G_\sigma)_{i,j}= \frac{\langle \Psi|\tilde c^\dagger _{j,\sigma} \tilde c _{i,\sigma}|\Psi \rangle}{\langle \Psi|\Psi \rangle}\,.
\end{equation}
It is straightforward to verify that $\tilde G_\sigma$ is diagonal. To match it with 
the occupancy of the target density matrix $\Lambda$, 
we have the following condition:
\begin{equation}
\lambda_i=\frac{\sum _{ \{n\}, \exists n=i }|d _{n_1}\cdots d_{n_{N_{\sigma}}}|^2}{\sum_{ \{n\}}| d_{n_1} \cdots d_{n_{N_{\sigma}}}|^2}\,,
\label{eq:cond-solve-for-F}
\end{equation}
where in the numerator the sum is restricted to sets of $\{ n\}$ which contain $i$. 
This gives $N_s$ equations which can be solved to determine the $N_s$ elements of the diagonal matrix $D$.
The resulting pseudo-BCS wave function $|\Psi\rangle$  reproduces the target density matrix:
\begin{equation}
\langle c^\dagger_{j,\uparrow}c_{i,\uparrow} \rangle =P \Lambda P^\dagger={\mathcal G}_\uparrow;
\ \ 
\langle c^\dagger_{j,\downarrow}c_{i,\downarrow} \rangle =Q \Lambda Q^\dagger={\mathcal G}_\downarrow\,.
\end{equation}

The pseudo-BCS wave function  $|\Psi\rangle$  in Eq.~(\ref{eq:BCS-expand-det}) has the characteristic matrix 
\begin{equation}
F=PDQ^T\,.
\end{equation}
Comparing it to Eq.~(\ref{eq:pairing-SVD}), we see that $U=P$ and $V=Q^\star$.
It is useful to consider the special case of AGP wave functions.
If the natural orbitals are real, as in RHF-type orbitals in quantum chemistry, we have  $P=Q$. 
Otherwise, if the orbitals are complex, $P$ should be equal to $Q^\star$ for BCS pairing. 
That is, 
we must 
organize the degenerate natural orbitals in $P$ and $Q$ such that 
$\uparrow$-spin and $\downarrow$-spin orbitals with the same occupancy are arranged 
in complex conjugate pairs. 
For example, for the attractive Hubbard model ($U/t < 0$ in Eq.~(\ref{eq:Ham-Hub})), the column-vectors 
$P_{i, n}$  and 
$Q_{j, n}$ should give 
a plane-wave with momentum ${\bf k}$ and ${\bf -k}$, respectively,
so as to have $|{\bf k}\rangle$ for $\uparrow$-spin and $|{\bf -k}\rangle$ for $\downarrow$-spin be coupled,
and the corresponding $F$ matrix in the form $PDP^\dagger$.
In the reverse direction, if we obtain $F$ via a BCS mean-field calculation, which yields diagonal matrix elements in momentum 
space given by $v_{\bf k}/u_{\bf k}$ 
\cite{Carlson_2011},  the SVD of $F$ will result in 
a form $UDU^\dagger$, consistent with the construction from 
the one-body density matrix discussed above.

In general, when a degeneracy is present in the density 
matrix eigenvalues in Eq.~(\ref{eq:target_DM}), there are extra 
degrees of freedom in how the eigenvectors (i.e., natural orbitals) among the degenerate 
set can be paired in constructing the pseudo-BCS wave function.
In the AGP case, this is determined by BCS theory as discussed above. With pseudo-BCS, we have adopted the approach of 
minimizing the variational energy. 

 
We next discuss the solution of Eq.~(\ref{eq:cond-solve-for-F}) for $D$. 
For wave functions in BCS form without number projection, that is, within grand canonical ensemble, the density matrix decomposition procedure 
in Eq.~(\ref{eq:cond-solve-for-F}) has the following 
exact solution  \cite{Shi_Many_2017}:
\begin{equation}
|d_i|=\sqrt{\lambda _{i}/(1-\lambda _{i})}\,.
\label{eq:appr-d}
\end{equation}
For our pseudo-BCS wave functions, this is an approximate solution which should become more accurate as the system becomes larger.
In our applications below, we use  Eq.~(\ref{eq:appr-d}) to determine the amplitude of $d_i$. 

Only the amplitude of $d_i$ is determined by 
Eq.~(\ref{eq:cond-solve-for-F}) or (\ref{eq:appr-d}), however. Any complex phase factor can be assigned 
to $d_i$ without affecting the density matrix $\tilde G$. We use this degree of freedom to minimize the variational energy of the resulting pseudo-BCS 
wave function. For the Hubbard Hamiltonian, this turns out to be equivalent to minimizing the  double occupancy 
$\langle \sum_i n_{i,\uparrow}n_{i,\downarrow}\rangle$, which can be done efficiently.
This is further discussed in Sec.~\ref{sec:level5}.
In general, one can also replace the solution of  Eq.~(\ref{eq:cond-solve-for-F}) with a minimization procedure, to find the set of complex diagonal elements $\{ d_n\}$ which
minimize the difference between the density matrix and the target, 
as well as the variational energy, of the resulting pseudo-BCS wave function. 

In our applications below, we sometimes deliberately break the spin symmetry in the Hubbard model, by applying a spin-dependent 
pinning field to induce antiferromagnetic correlations \cite{Qin_Coupling_2016}. Similar to the situation in UHF, this preserves the  condition that 
${\mathcal G}_\uparrow$ and ${\mathcal G}_\downarrow$  
share the same eigenvalues but have different eigenvectors (orbitals). For systems that do not satisfy this condition, one could still enforce the condition and adopt the approach of
seeking a pseudo-BCS wave function whose density matrices are closest to the target, but that would of course be an additional level of approximation.

\section{\label{sec:level3}Self-consistent AFQMC with pseudo-BCS wave function}

As mentioned earlier, the pseudo-BCS wave function can provide a better ansatz not only for systems with superconducting correlations, but for 
systems with purely repulsive electron-electron interactions. By 
generalizing the concept of AGP, the pseudo-BCS wave function allows paired states in the sense of UHF orbitals. For 
many systems with antiferromagnetic correlations, the UHF provides a better description (at the cost of symmetry breaking). 
In such systems, the pseudo-BCS with UHF orbitals
is analogous to AGP with RHF-like orbitals in BCS systems. 

In this section, we illustrate one application of the pseudo-BCS form by coupling it to the AFQMC framework.
We input into the AFQMC 
a trial wave function chosen in the form of a pseudo-BCS wave function, and then  self-consistently improve it with the density matrix from AFQMC, using the 
decomposition discussed in the previous section. 
Below we  
first briefly review the concept and algorithms of AFQMC, before
introducing the procedure for realizing the self-consistency with the pseudo-BCS wave function.

\subsection{\label{sec:level31}AFQMC and self-consistency}

The ground-state AFQMC method relies on  imaginary time evolution from an initial Slater determinant $|\Phi_I\rangle$
(or any linear combination of Slater determinants): 
\begin{equation}
|\Psi_{0}\rangle=e^{-\tau H}|\Phi_I\rangle=e^{-\tau (K+V)}|\Phi_I\rangle\,,
\end{equation}
which will project to the ground state of $H$ if the overlap $\langle\Psi_{0}|\Phi_I\rangle$ is non-zero. To 
realize 
this imaginary time evolution, we first use Suzuki-Trotter decomposition 
to break up the imaginary-time evolution operator 
\begin{equation}
e^{-\tau H }  
\approx   \prod ^{N}_{n=1} e^{-\delta\tau K/2}\,e^{-\delta\tau V}\,e^{-\delta\tau K/2}\,,
\end{equation}
where $\delta\tau=\tau/N$.
We then apply the Hubbard-Stratonovich transformation to the two-body term;
\begin{equation}
e^{-\delta\tau V}\!\!=\!\! \int\mathrm{d}\textbf{x}\,p(\textbf{x})\,e^{\hat{h}(\textbf{x})}\,,
\end{equation}
where $\hat{h}(\textbf{x})$ is a one-body operator which dependents on the auxiliary-field vector $\textbf{x}$, $p(\textbf{x})$ is a probability density function, and $e^{\hat{h}(\textbf{x})}$ propagate a Slater determinant $|\Phi\rangle$ to another Slater determinant  $|\Phi'\rangle$. 
Putting these together \cite{Zhang_Constrained_1997}, we have
\begin{equation}
\begin{aligned}
e^{-\tau H} \!\! &= \prod ^{N}_{n=1}e^{-\delta\tau (K+V)}\!\! \\
&\approx   \prod ^{N}_{n=1}\!\int \mathrm{d}\textbf{x}_n\,e^{-\delta\tau K/2}p(\textbf{x}_n)e^{\hat{h}(\textbf{x}_n)}e^{-\delta\tau K/2}\,.
\end{aligned}
\end{equation}
The integrals on the right-hand side are in  many dimensions and will require Monte Carlo. 
 In a sense, the imaginary-time evolution in AFQMC can be viewed as an ensemble of random walks in a space of Slater determinants.
 This population of Slater determinants in the random walks are each orthonormal, but not orthogonal to each other, as the branching random walks 
occur in an over-complete determinant space \cite{lecturenotes-2019}.

The anti-symmetry of fermions will cause an arbitrary sign or phase to develop 
in each of the Slater determinants during the stochastic propagation. 
If a walker propagates to become perpendicular to the ground state, this walker will effectively cease to 
contribute 
under further projection. 
The number of such  walkers will in general grow exponentially with projection time, which will result in large statistical fluctuations. 
Eliminating them is an exact condition which removes the sign or phase problem \cite{Zhang_Constrained_1997,Zhang-Krakauer-2003-PRL}; 
in practice the condition is implemented using a 
trial wave function $|\Psi_T\rangle$, which is applied to constrain the random walk paths.
In Hubbard-like models, this constraint is  
$\langle \Psi_T|\Phi\rangle > 0$ 
for ground-state calculations, but for a more general form of interaction a condition involving the phase is needed \cite{Zhang-Krakauer-2003-PRL}.
If the trial wave function is the exact ground state, then this constraint is unbiased. 
Many studies have shown that even a free-electron or HF wave function typically gives high accuracy 
in a variety of correlated systems \cite{CPMCModel,CPMCModel2,CPMCMaterial}.

To reduce the dependence on the trial wave function in the constrained path or phaseless approximation, 
a self-consistent procedure was introduced based on Slater-determinant-type of wave functions  \cite{Qin_Coupling_2016}.
Here we introduce a new self-consistent AFQMC based on the pseudo-BCS decomposition. 
A trial wave function in the form of a pseudo-BCS is used as a constraint to carry out 
the AFQMC calculation, from which a density matrix is computed.  We then apply the density matrix decomposition to the result 
to obtain a new pseudo-BCS wave function, which is used  as the trial wave function in the next iteration of AFQMC calculation. 
The procedure 
is repeated until the density matrix or other physical observables computed from AFQMC are converged. 
The approach proposed here achieves self-consistency via the one-body density 
matrix without needing to involve a fictitious mean-field calculation \cite{Qin_Coupling_2016,CPMCModel2}.

\subsection{\label{sec:level33}Additional details: application of pseudo-BCS trial wave function  in AFQMC}
We provide some of the formalism and details \cite{Carlson_2011,Shi_Many_2017} 
necessary 
to apply a pseudo-BCS trial wave function in AFQMC and to realize the self-consistent procedure described above.
%
We define the overlap matrix  between the pseudo-BCS wave function 
of Eq.~(\ref{eq:PseudoBCS}) 
and a Slater determinant of  Eq.~(\ref{eq:sd}) as:
\begin{equation}
A\equiv \Phi^T_{\uparrow}\cdot F^\star\cdot \Phi _{\downarrow}\,.
\end{equation}

The overlap between $|\Psi\rangle$ and $|\Phi\rangle$ is then 
\begin{equation}
\langle \Psi |\Phi \rangle =(-1)^{N_{\sigma}(N_{\sigma}-1)/2} 
N_{\sigma}!\det(A)\,.
\end{equation}
(The global sign and coefficient above will have no affect
in the calculations  in this paper.)

The one-body term mixed-estimate \cite{lecturenotes-2019} is
\begin{equation}
\label{eq:Gup-mixed-BCS-det}
\frac{\langle \Psi|c^\dagger_{i,\uparrow}c _{j,\uparrow}|\Phi \rangle}{\langle \Psi|\Phi \rangle}=[F^\star\cdot \Phi_{\downarrow}\cdot A^{-1}\cdot \Phi^T_{\uparrow}]_{i,j}\,,
\end{equation}
and
\begin{equation}
\label{eq:Gdn-mixed-BCS-det}
\frac{\langle \Psi |c^\dagger_{i,\downarrow}c _{j,\downarrow}|\Phi \rangle}{\langle \Psi |\Phi \rangle}=[F^\dagger \cdot \Phi_{\uparrow}\cdot (A^{-1})^T \cdot \Phi ^T_{\downarrow}]_{i,j}.
\end{equation}
In order to evaluate the pure estimator (as opposed to the mixed estimator) of the density matrix or other observables, back-propagation is 
needed \cite{Zhang_Constrained_1997,Wirawan-PRE}. The back-propagation of a pseudo-BCS wave function is more subtle, and a 
scheme to ensure numerical stability has recently been proposed \cite{Vitali_Calculating_2019}, which we adopt here. 

\section{\label{sec:level4}Results}

In this section, we use the Hubbard model as an example to illustrate the  method described above and
show the improvement of the self-consistent AFQMC with a pseudo-BCS form of the trial wave function. 
In Sec.~\ref{sec:level41} we show results in the pure Hubbard model with $t'=0$. Here extensive results exist from previous studies which 
have shown that AFQMC with the usual Slater determinant trial wave functions is quite accurate, and 
we use this case as a benchmark. Then in Sec.~\ref{sec:level42} we apply the method to the case with  $t' \ne 0$ where it is shown that our
pseudo-BCS leads to an improvement in parameter regimes where the single determinant trial wave function is less accurate.

\subsection{\label{sec:level41}Illustration in the pure Hubbard model}

We first study the Hubbard model with only nearest-neighbor hopping, i.e., with $t_{i,j}=t$ for near-neighbors $\langle i,j\rangle$ and $t_{i,j}=0$ otherwise.
We will work with a $4\times4$ lattice under periodic boundary condition as a test case, in which exact diagonalization (ED) can be performed 
straightforwardly. We choose $U=4t$, 
and doping $h\equiv 1-(N_\uparrow+N_\downarrow)/N_s=1/8$.
This doping represents the parameter regime where the sign problem is the most severe, and the constraint error is the largest because of both open-shell
and interaction effects.
The ground state of this system has three fold degeneracy, distinguished by their overall momentum, $(0,0)$ and $(0,\pi)$ or $(\pi,0)$ \cite{Parola_d_1991}. 
The $(0,0)$ case has a degeneracy in the occupancy of the natural orbitals.
We focus on the latter situation below in order to compare with single-determinant (SD)
cases with no ambiguity. 

\begin{figure}[htbp]
\centering
\includegraphics[scale=0.225]{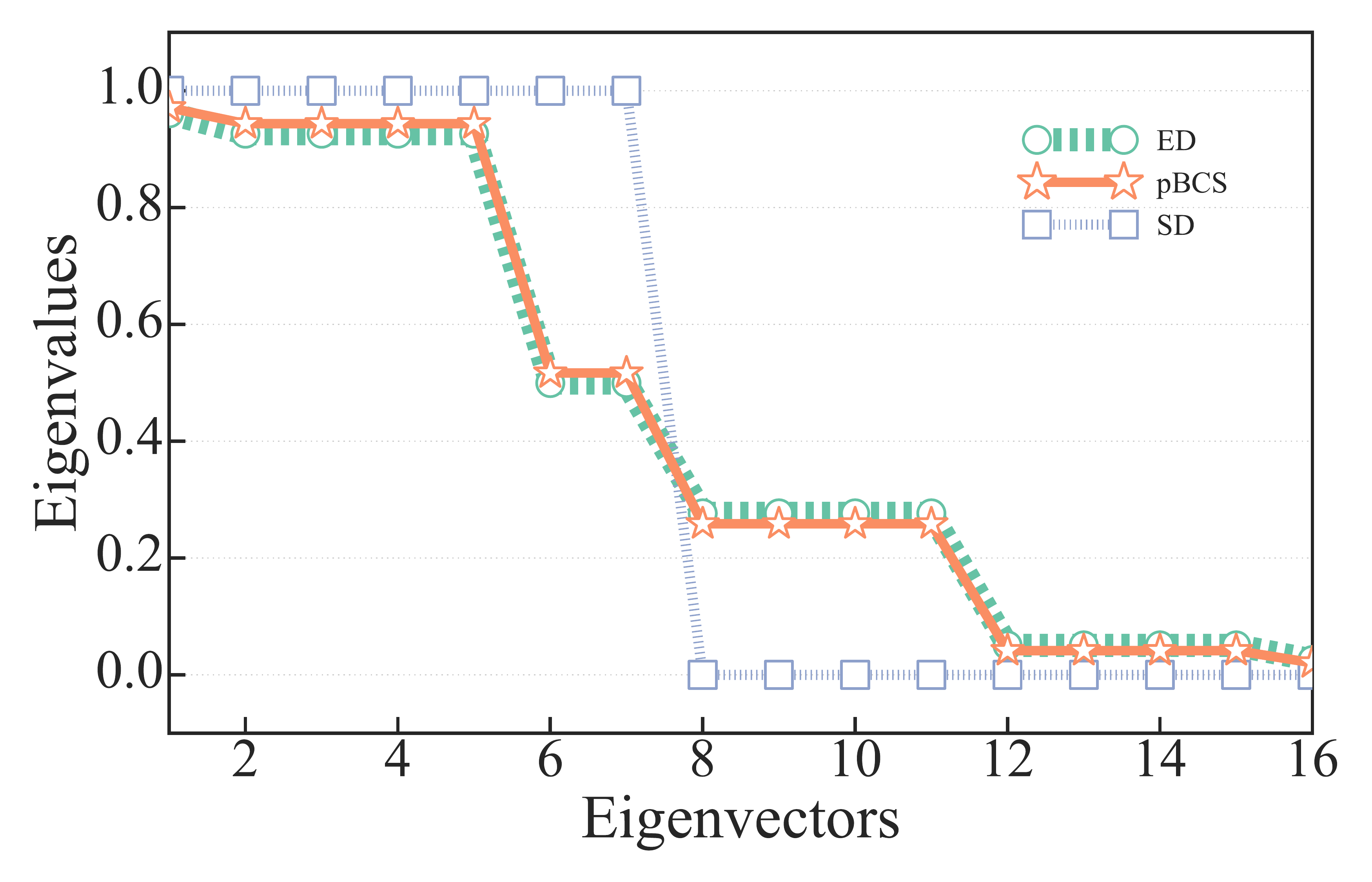}
\caption{
Illustration of the density matrix decomposition to construct the pseudo-BCS wave function. 
The system is a $4\times4$ Hubbard model, 
with $N_\uparrow=N_\downarrow=7$ (i.e., $1/8$-doping), 
$U=4t$, with periodic boundary condition 
and overall momentum $(0,\pi)$ or $(\pi,0)$. ``SD" indicates natural orbital, i.e., the result of a single Slater determinant constructed from the eigenvectors of the density matrix. 
``pBCS" indicates the pseudo-BCS wave function constructed from the procedure in Sec.~\ref{sec:Algorithm}. 
}
\label{fig:1}
\end{figure}

In Fig.~\ref{fig:1}, we illustrate the density matrix decomposition procedure discussed in Sec.~\ref{sec:Algorithm}. 
The pseudo-BCS wave function is generated 
from the exact one-body density matrix from ED. 
The resulting eigenvalues of the pseudo-BCS wave function from the decomposition is compared with the exact results. The result of a single 
Slater determinant wave function constructed from natural orbitals 
(which is equivalent to the free-electron wave function in this case)
 is also shown.

\begin{figure}[htbp]
\includegraphics[scale = 0.1]{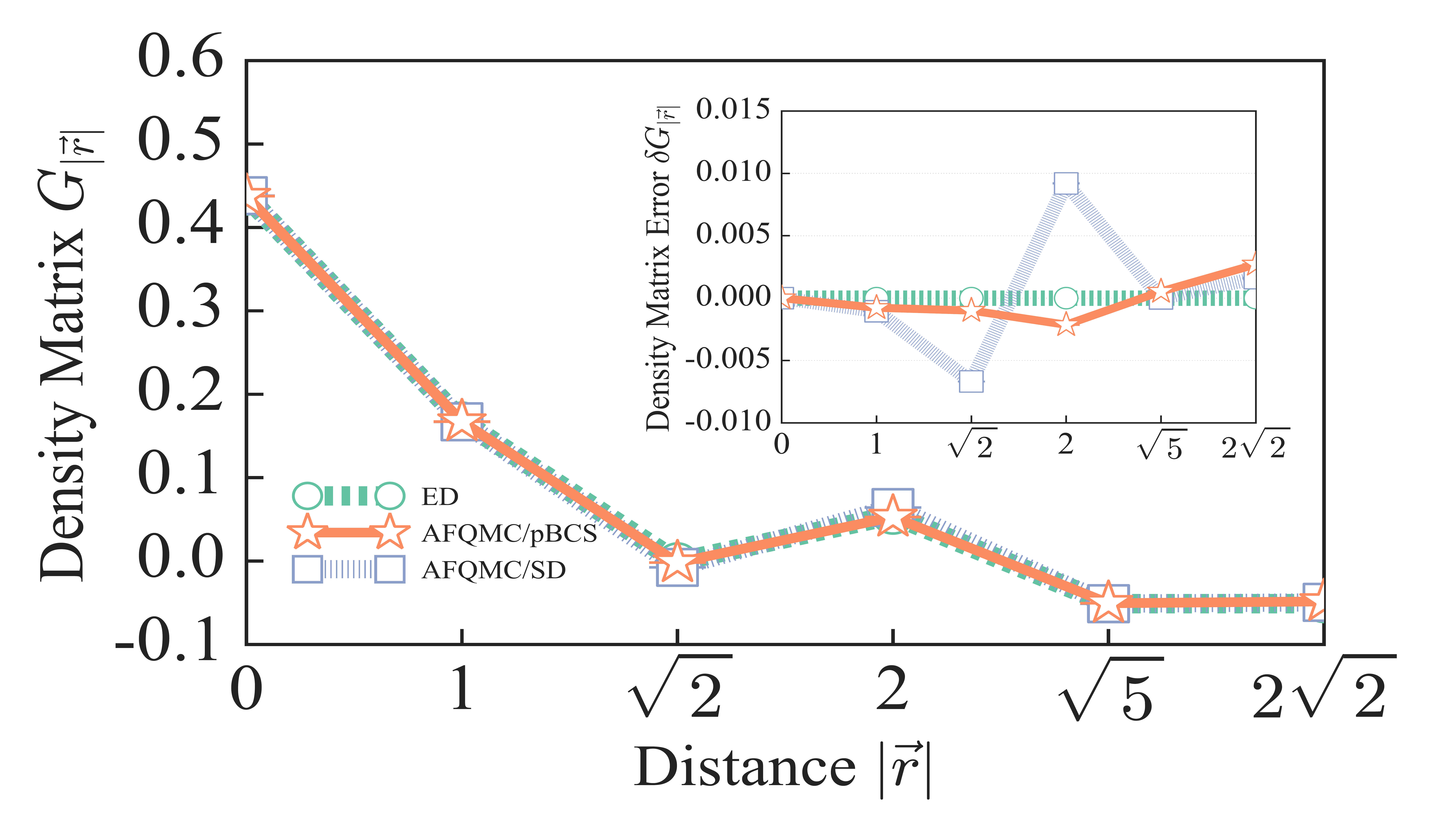}   
\caption{The density matrix $G_{i,j}$ computed from AFQMC, shown as a function of the distance $\vec{r}$ between sites $i$ and $j$ under PBC, compared 
with the exact result from ED. The inset shows the 
 error $\delta G_{i,j}$ with respect to the ED result. 
The system and setup are the same as in Fig.~\ref{fig:1}.
Statistical error bars are shown but are smaller than symbol size.
}
\label{Fig.ExactInput}
\end{figure}

For this test, we obtain  the exact ground state density matrix from ED, and 
apply the density matrix decomposition to obtain a pseudo-BCS wave function.
This wave function is then applied in 
AFQMC as the trial wave function.
%
In Fig.~\ref{Fig.ExactInput}, the density matrix computed from AFQMC using this trial wave function is compared with that from ED. 
For reference, the density matrix computed from AFQMC using a 
SD 
trial wave function which is formed with the 
natural orbitals with the $N_\sigma$ largest engenvalues (occupancy) is also shown.
We see that the results from AFQMC/pseudo-BCS are not exact, despite using the exact density matrix to generate the pseudo-BCS. 
This is not surprising, because the pseudo-BCS wave function is not the exact
many-body wave function, and apparently still incurs a finite constraint error in AFQMC.

\begin{center}
\begin{table*}[]
\caption{Comparison of the computed total energies from self-consistent AFQMC using trial wave functions of pseudo-BCS form
(SC AFQMC/pBCS) and single Slater determinant (SC AFQMC/SD), as well as one-shot AFQMC using free-electron 
trial wave functions  (AFQMC/FE) for various systems,
compared to DMRG
in $4\times 8$ and $4\times 16$ cylinders with pinning fields applied at the edges. 
All systems have $U=4t$, and the details of the pinning fields in the cylindrical systems are given in the text. 
}
\begin{tabular}{ccccccccccccccc}
\hline \hline
$L_x\times L_y$ & & $N_\uparrow\ \ N_\downarrow$ & & $t'/t$ &  & AFQMC/FE&  & SC AFQMC/SD &  & SC AFQMC/pBCS &  & ED/DMRG &  \\\hline
$4\times8$ & &$16\uparrow16\downarrow$& & $0.3$ &  & $-27.271(1)$ &  & $-27.611(1)$ &  & $-27.682(1)$ &  & $-27.6924$ &  \\
$4\times8$ & &$16\uparrow16\downarrow$& & $0.35$ &  & $-27.802(1)$ &  & $-27.805(1)$ &  & $-27.914(1)$ &  & $-27.9755$ &  \\
$4\times16$ & &$32\uparrow32\downarrow$& & $0.3$ &  & $-56.110(4)$ &  & $-56.268(3)$ &  & $-56.176(3)$ &  & $-56.236$ &  \\
$4\times16$ & &$32\uparrow32\downarrow$& & $0.4$ &  & $-57.801(3)$ &  & $-57.943(2)$ &  & $-57.868(3)$ &  & $-57.930$ &  \\
\hline \hline
\end{tabular}
\label{Table.Energy}
\end{table*}
\end{center}

\subsection{\label{sec:level42}Hubbard model at half-filling with $t'$}

In this section 
we study the Hubbard model at half-filling with both near-neighbor ($t$) and next-nearest-neighbor hopping ($t'$).
These systems provide a good test case for us, with the presence of different magnetic orders  \cite{Hubbardt-prime1987,AFM_Hubbard_tp_Mizusaki_2006,AFM_Hubbard_tp_Nevidomskyy_2008,Sorella_bf_BCS_tp_2008}.
We will focus on its magnetic correlations, 
applying pinning fields  \cite{Pinning1, Pinning2,Pinning3}  under cylindrical boundary conditions, 
i.e., periodic along $x$-direction and open along $y$.
The pinning fields are applied at the edges of the cylinder, adding a one-body external potential term  
$\sum_{i, \sigma} u_{i,\sigma} n_{i,\sigma}$
to the Hamiltonian of Eq.~(\ref{eq:Ham-Hub}), with
 $u_{i,\uparrow}=-u_{i,\downarrow}=(-1)^{i_x} u_0$ 
  for $i_y=1$ and $L_y$, 
  and $u_{i,\sigma}=0$ for all other sites.
 The  strength of the pinning field is fixed at $u_0=0.25 t$ in our calculations.
 The cylindrical boundary condition and pinning fields break translational symmetry along the $y$-direction and induces AFM 
 correlations.
 Under this setting, two-body spin correlation functions in periodic systems can be probed by one-body spin densities:
$S^z_{i}=\langle (n_{i,\uparrow}-n_{i,\downarrow})/2\rangle$.
 It provides a convenient way to detect the presence and nature of long-range AFM orders including collective modes such as stripes 
 \cite{stripe-science,Qin_Coupling_2016}.
 We will study width-4 cylinders which can be treated very accurately by density matrix renormalization 
 group (DMRG) calculations \cite{PhysRevLett.69.2863}, which we perform using the  ITensor Library\cite{ITensor} and with which we benchmark our self-consistent 
 AFQMC results.

\begin{figure}[htbp]
\includegraphics[scale = 0.225]{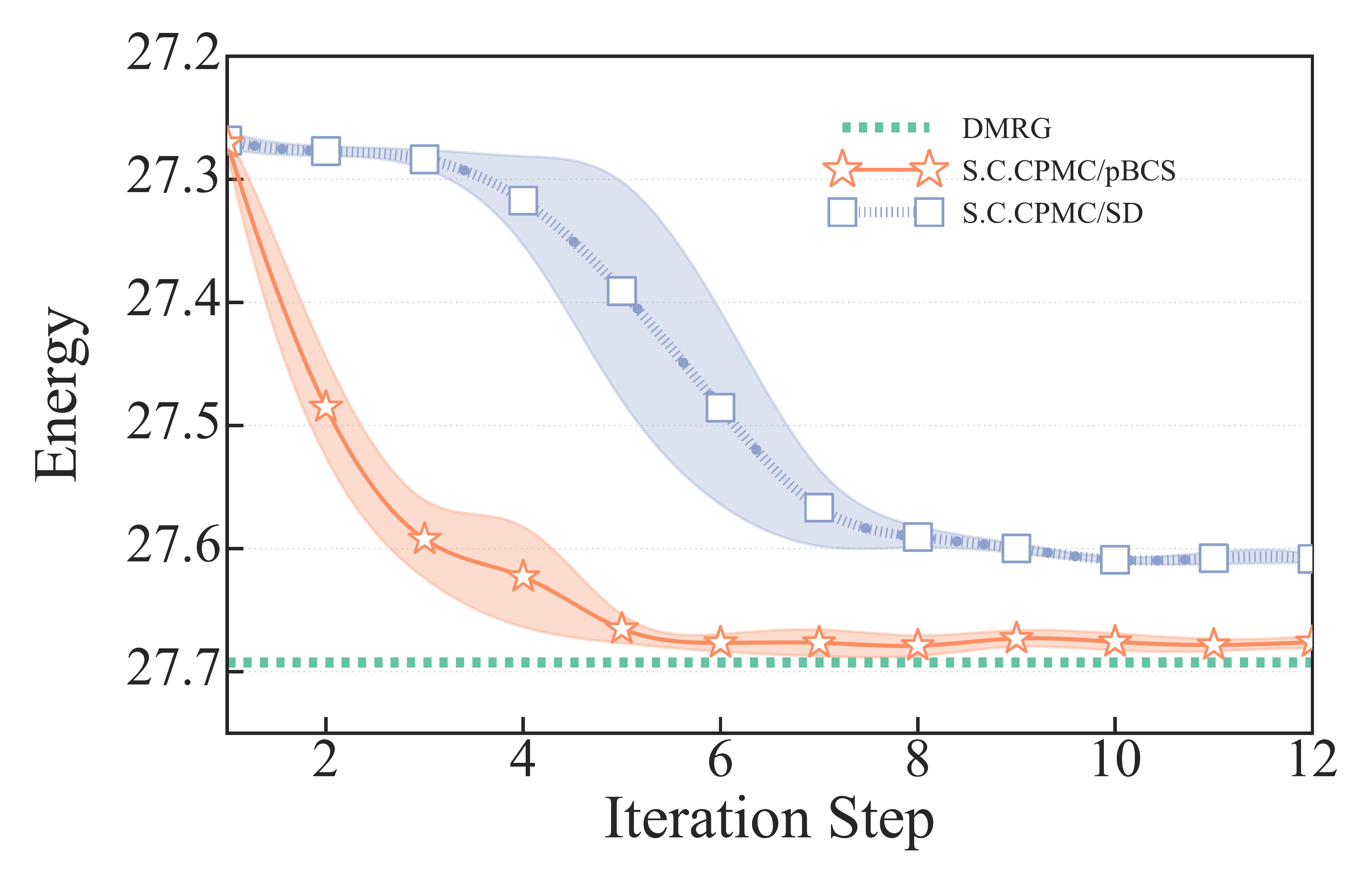}   
\includegraphics[scale = 0.225]{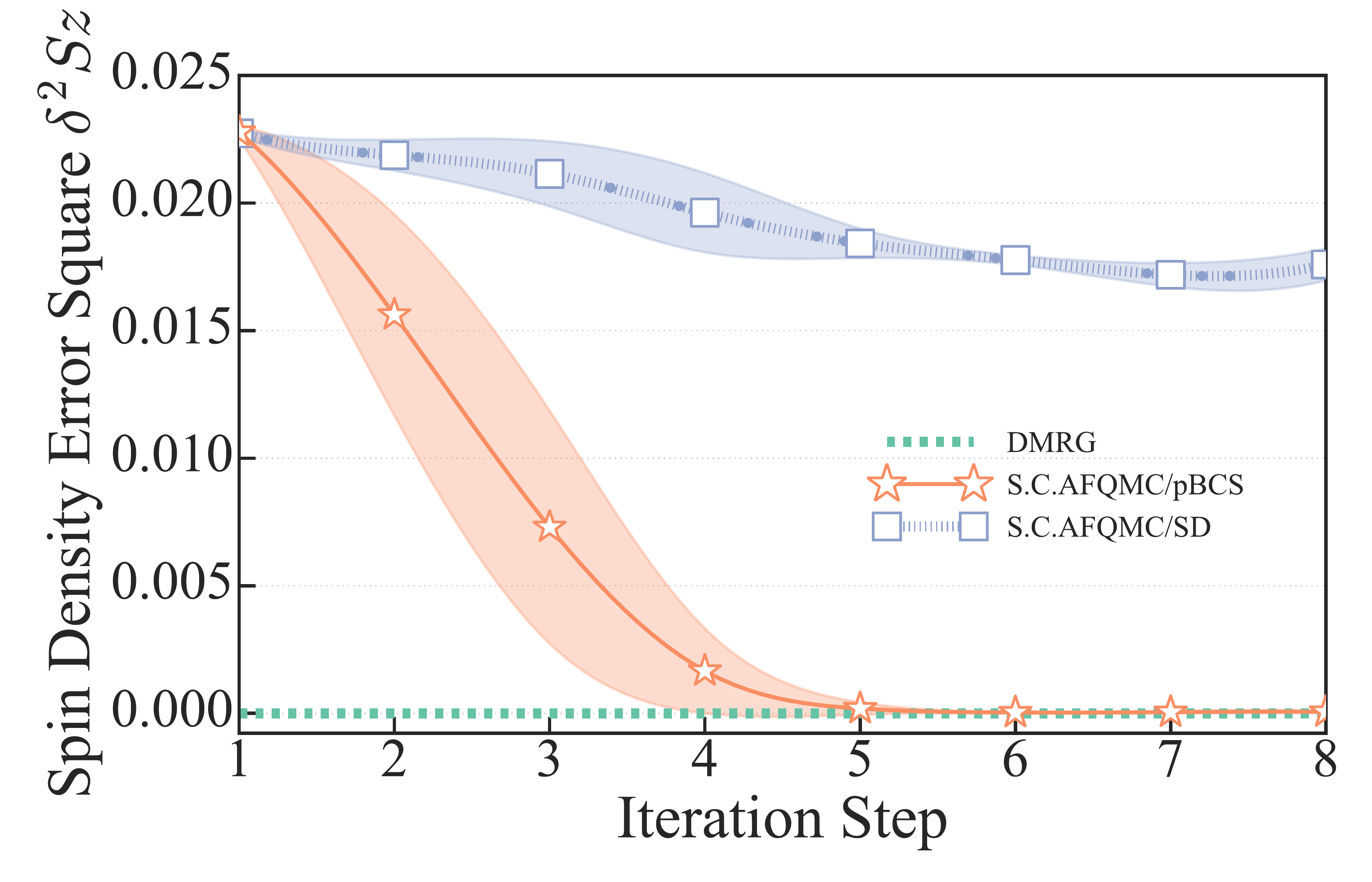}   
\caption{Convergence of the self-consistent AFQMC calculation using a trial wave function of the pseudo-BCS form versus a single Slater determinant
from natural orbitals.
The system is a $4\times8$ Hubbard cylinder at half-filling, with pinning field applied on both edges, with $U=4t$ and $t'=0.3t$.
The top panel shows the ground-state energy and the bottom panel shows the mean squared error $\delta_{S^z}^2$ in the spin density with respect to DMRG. 
The SC procedures were repeated multiple times with different random number seeds to estimate the uncertainties in the convergence process, 
with the standard deviations shown by the shading on the curve. 
} 
\label{Fig.SCAFQMC}  
\end{figure}

We carry out
self-consistent AFQMC calculations and benchmark the computed ground-state energy and  spin densities. 
For the SC AFQMC/SD calculations, we keep the trial wave function in the form of a single Slater determinant, which is obtained in
the self-consistent iteration from the natural orbitals 
of the computed density matrix, taking the $N_\sigma$ leading natural orbitals with the largest occupancies \cite{Qin_Coupling_2016}. 
For the SC AFQMC/pBCS, we compute 
the one-body density matrix with back-propagation  \cite{Vitali_Calculating_2019} and then apply our density matrix decomposition procedure 
discussed in Sec.~\ref{sec:Algorithm}. 
%
In Fig.~\ref{Fig.SCAFQMC} we show the convergence of the ground-state energy and spin density as a function of self-consistency iterations. 
For the spin density, we measure the mean squared deviation
 $\delta_{S^z}^2 \equiv \sum^{N_s}_i (S^z_{i}-S^z_{i,{\rm DMRG}})^2/N_s$ with respect to the DMRG reference result.
Both self-consistency processes were initialized using 
the free-electron wave function, which provides the wrong initial input in most cases 
as it
is not magnetically ordered. 
Both sets of SC calculations yield improved results over the initial AFQMC/FE result, as expected.
The SC AFQMC/pBCS shows both a faster convergence and better converged results over the  SC AFQMC/SD.


The ground-state energies are listed in Table~\ref{Table.Energy} for the system above and several other cylindrical systems, using the two different self-consistent approaches, as well as
one-shot AFQMC calculations with the free-electron trial wave function. 
The final energy for self-consistent AFQMC with pseudo-BCS is calculated using 
the pseudo-BCS trial wave function with lowest variational energy after density matrix convergence. 
The free electron trial wave function is degenerate in $4\times16$ with $t'/t=0.3$. We break the degeneracy by 
 solving the non-interacting 
Hamiltonian with a small twisted angle $(0.01,0.01)$ applied in the boundary condition.
Consistent with the trend observed in Fig.~\ref{Fig.SCAFQMC},
both SC procedures are  seen to improve the energy, with 
the SC AFQMC/pBCS giving systematic errors of $\approx 0.2$\% or  less compared to DMRG.

\begin{figure*}[htbp]
\centering
\includegraphics[scale = 0.25]{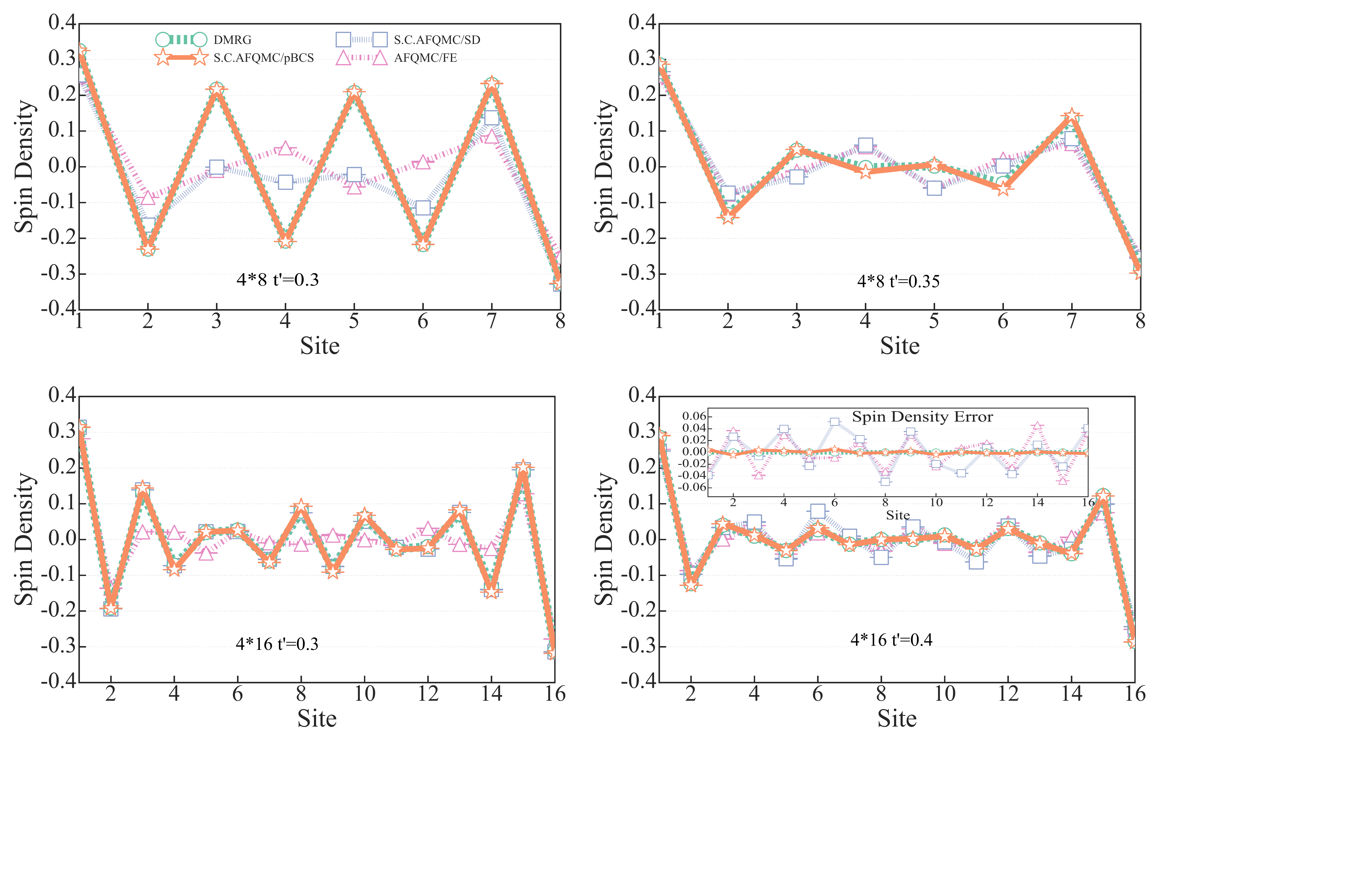}
\caption{Resolving different magnetic orders.
The computed spin densities  with self-consistent AFQMC, using either pseudo-BCS (SC AFQMC/pBCS)  or single determinant
(SC AFQMC/SD) form of trial wave functions, 
 are shown in four systems with different magnetic correlations, together with the one-shot
AFQMC/FE and reference DMRG results for comparison. 
All systems have  $U=4t$ and are at half-filling, with cylindrical boundary condition and edge pinning fields applied; the system size and next-near-neighbor hopping value $t'$ are indicated in each panel. All four panels use the same symbols as indicated in the top left panel.  The inset in the lower 
right panel shows a magnified view of the middle region. 
The spin density is plotted versus site label along a cut in the $y$-direction of the cylinder, with $i_x  = 2$. 
For all systems, the results are statistically indistinguishable (after accounting for the alternating sign)
 with respect to $i_x$.  
}
\label{Fig.Tmodel}
\end{figure*}

In Fig.~\ref{Fig.Tmodel}, we study the spin density more systematically. 
As a function of the next near-neighbor hopping amplitude $t'$, different magnetic correlations arise, which provides an excellent
test ground for the self-consistency procedure via pseudo-BCS decomposition.
The converged spin densities
from  self-consistent AFQMC with pseudo-BCS decomposition are shown and compared to the exact results from DMRG. Results from 
single determinant trial wave functions are also presented, including both the one-shot calculation using the free-electron trial wave function and the 
self-consistent constraint from truncated natural orbitals. 
 We see that the SC AFQMC with pseudo-BCS is able to resolve the details of the spin 
order, yielding results in excellent agreement with DMRG.


\section{\label{sec:level5}DISCUSSION}

\subsection{\label{sec:level51}Variational optimization of the phases of $\{d_n\}$}

As mentioned in Sec.~\ref{sec:level23}, the density matrix decomposition determines only the absolute values of $\{d_i\}$, allowing for arbitrary phase factors. We 
adjust the $(N_\sigma-1)$ independent phase factors to minimize the variational energy of the pseudo-BCS wave function
\begin{equation}
\label{eq:varE-BCS}
E_\Psi=\frac{\langle \Psi|H|\Psi\rangle}{\langle \Psi|\Psi\rangle}=\frac{\langle \Psi|K+V|\Psi\rangle}{\langle \Psi|\Psi\rangle}\,.
\end{equation}
The variational energy of a pseudo-BCS wave function can be evaluated by a Monte Carlo sampling of the ket, which we have used in earlier 
studies  \cite{Carlson_2011,Vitali_Calculating_2019} and which we briefly describe below.

To evaluate the expectation value of an observable with respect to a pseudo-BCS (or AGP) wave function, we can use the expansion in 
Eq.~(\ref{eq:BCS-expand-det}) to write the ket $|\Psi\rangle$ as a linear combination of Slater determinant, 
$|\Psi\rangle=\sum_\Phi c_\Phi |\Phi\rangle\,$,
where $c_\Phi=d_{n_1}...d_{n_{N_{\sigma}}}$ and $|\Phi\rangle$ is the Slater determinant obtained by occupying the set of natural orbital pairs specified by
the indices $\{n_1, n_2, \cdots, n_{N_{\sigma}}\}$. 
Equation~(\ref{eq:varE-BCS}) can then be written as
\begin{equation}
\label{eq:var-BCS-expand-det}
E_\Psi=\frac{\sum_\Phi \frac{\langle \Psi|H|\Phi\rangle}{\langle \Psi|\Phi\rangle}\,c_\Phi \langle \Psi|\Phi\rangle}{\sum_\Phi c_\Phi \langle \Psi|\Phi\rangle}
=\frac{\sum_\Phi \frac{\langle \Psi|H|\Phi\rangle}{\langle \Psi|\Phi\rangle}\,|c_\Phi|^2}{\sum_\Phi |c_\Phi|^2}\,,
\end{equation}
which is similar to 
the ``mixed estimator'' in AFQMC. 
The sum over $\Phi$ 
contains a combinatorial number of terms, which makes an explicit summation impractical for larger systems. 
We can sample the Slater determinants $|\Phi\rangle$ according to  $|c_\Phi^2|$ by,  for example, 
a Markov chain Monte Carlo procedure in the discrete space of indices $\{n\}$, proposing to swap one of the occupied index $n_i$ with one from 
the unoccupied set, $n_i'$, 
and 
accepting the move  based on $|d_{n_i'}/d_{n_i}|^2$. 
The mixed estimate between $\langle \Psi|$ and $|\Phi\rangle$ in the numerator can be computed 
using Green functions as shown in Eqs.~(\ref{eq:Gup-mixed-BCS-det}) and (\ref{eq:Gdn-mixed-BCS-det}), and applying Wick's theorem 
for the two-body term in $V$ as needed \cite{Vitali_Calculating_2019}. 
(
The one-body and two-body terms of a pseudo-BCS wave function could also be computed using 
related characteristic polynomial \cite{AGP_dm_Khamoshi_2019}.)

The phases of $\{d_i\}$ only enter in the local energy  in Eq.~(\ref{eq:var-BCS-expand-det}). They do not affect the Monte Carlo sampling or the sampled $|\Phi\rangle$'s. The phases in $|\Psi\rangle$  can be optimized using, for example,  standard variational minimization techniques employed in 
variational Monte Carlo \cite{Sandro_VMC}. In the Hubbard model, the kinetic term $K$ is invariant 
with respect to the phase; the $V$ term easily decouples into single particle forms for the two spin sectors. 

\subsection{\label{sec:level52}Generalization of the pseudo-BCS form}

 
We briefly comment on generalization of the pseudo-BCS approach beyond systems we have discussed, which 
has been restricted to Hamiltonians with no spin-flip terms and singlet pairing with spin balance. 

With a  one-body term $K$ that contains spin-flip terms, such as systems with spin-orbit coupling (SOC),
the Slater determinant state $|\Phi\rangle$ contains spin-orbitals 
and is described by a  $2N_{s} \times (N_\uparrow+N_\downarrow)$ matrix \cite{Peter-Hao-Zhang-SOC-J-Phys-Chem-Solids-2019}: 
$\binom{\Phi_{\uparrow}}{\Phi_{\downarrow}}$.
The corresponding one-body density matrix is a 
$2N_{s} \times 2N_{s}$ matrix.  
The eigen decomposition of the density matrix gives $2 N_s$ natural spin-orbitals $U$.
The pseudo-BCS wave function is given by a matrix $F$, in the form of: 
$F=U D U^T$,
where $D$ is a skew-symmetric matrix. 
Similar to the spin-decoupled case, the elements of $D$, $\{ d_{i,j} \}$, are 
to be determined by the eigenvalues $\{\lambda\}$ of the density matrix, where a pair of comparable eigenvalues $\lambda_i=\lambda_j$
gives $d_{i,j}=-d_{j,i}$. 

The equations for overlap and the mixed estimator  between a general Slater determinant $\Phi$ with spin-orbitals
and a pseudo-BCS wave function $\Psi $ from the above are given by: 
\begin{equation}
\langle \Psi|\Phi \rangle= \mathrm{pf}[\Phi ^T\cdot F^\star\cdot \Phi],
\end{equation}
and
\begin{equation}
\frac{\langle \Psi |c^\dagger _{i}c_{j}|\Phi \rangle}{\langle \Psi|\Phi \rangle}=[F^\dagger \cdot \Phi \cdot (\Phi^T F^\dagger \Phi)^{-1}\cdot \Phi ^T]_{i,j}\,,
\end{equation}
where $i$ (and $j$) is a general spin-orbital index (i.e., combining both the site index $i$ and spin $\sigma$ in the Hubbard Hamiltonian), and  
 $\mathrm{pf}$ denotes Pfaffian\cite{Pfaffian}.

\section{\label{sec:level6}SUMMARY}
In summary, we introduced an approach to use the one-body density matrix to build pseudo-BCS wave functions for a many-body systems, i.e., 
wave functions with a similar form to number-projected BCS.
They can be thought of as a generalization of BCS to systems with repulsive interactions such as in molecules and real materials.
Such wave functions can provide a better mean-field ansatz than any single Slater determinants, and 
a natural and very powerful extension 
beyond UHF.
 The pseudo-BCS wave functions are more versatile and has more variational freedom than the usual number-projected BCS
wave functions. 
They can be manipulated with costs similar to AGP or Slater determinants, and can be
used as the reference state in a variety of many-body methods.
By coupling pseudo-BCS wave functions to 
AFQMC self-consistently via a density matrix decomposition approach, we achieve 
a self-consistent AFQMC method, which yields 
improved results in the Hubbard model over state-of-the-art self-consistent AFQMC calculations based on  single Slater determinants.


\section{\label{sec:level7}ACKNOWLEDGEMENT}
We thank Ettore Vitali, Mingpu Qin, Yuan-Yao He for valuable discussion. This work is supported by 
the Simons Foundation 
and the many-electron collaboration. Computing was carried out at the computational facilities at William $\&$ Mary and the 
Flatiron Institute. The Flatiron Institute is a division of the Simons Foundation.

\bibliography{cite} 
\end{document}